# A fast algorithm for determining the linear complexity of periodic sequences

Jianqin Zhou

(Dept. of Computer Science, Anhui University of Technology, Ma'anshan 243002, P. R. China)

(E-mail: zhou63@ahut.edu.cn)

**Abstract:** A fast algorithm is presented for determining the linear complexity and the minimal polynomial of periodic sequences over GF(q) with period $q^n p^m$, where p is a prime, q is a prime and a primitive root modulo $p^2$. The algorithm presented here generalizes both the algorithm in [4] where the period of a sequence over GF(q) is $p^m$ and the algorithm in [5] where the period of a binary sequence is $2^n p^m$. When m=0, the algorithm simplifies the generalized Games-Chan algorithm.

　　The concept of linear complexity is very useful in the study of the security of stream ciphers for cryptographic applications. A necessary condition for the security of a key stream generator is that it produces a sequence with large linear complexity. Games-Chan algorithm in [1] was proposed to compute the linear complexity of sequences over GF(2) with period $2^n$, and was generalized to the sequences over GF($p^m$) with period $p^n$, where p is a prime, by Ding, Xiao and Shan in [2]. Wei, Xiao and Chen in [4] presented an algorithm to compute the linear complexity of sequences over GF(q) with period $p^n$, where p is a prime, q is a prime and a primitive root modulo $p^2$. They in [5] presented an algorithm to compute the linear complexity of sequences over GF(2) with period $2^n p^m$, where 2 is a primitive root modulo $p^2$.

　　In this paper, a fast algorithm is presented for determining the linear complexity and the minimal polynomial of periodic sequences over GF(q) with period $q^n p^m$, where p is a prime, q is a prime and a primitive root modulo $p^2$. The algorithm presented here generalizes both the algorithm in [4] where the period of a sequence over GF(q) is $p^m$ and the algorithm in [5] where the period of a binary sequence is $2^n p^m$. When m=0, the algorithm simplifies the generalized Games-Chan algorithm.

## 1. Preliminaries

　　We will consider sequences over GF(q). Let $x=(x_1,x_2,\cdots,x_n)$ and $y=(y_1,y_2,\cdots,y_n)$ be vectors over GF(q). Then define $x+y=(x_1+y_1,x_2+y_2,\cdots,x_n+y_n)$.

　　The generated function of a sequence $s=\{s_0, s_1, s_2, s_3, \cdots\}$ is defined by $s(x)= s_0 + s_1 x + s_2 x^2 + s_3 x^3 + \cdots = \sum_{i=0}^{\infty} s_i x^i$.

　　The generated function of a finite sequence $s^N = \{s_0, s_1, s_2, \cdots, s_{N-1}\}$ is defined by $s^N(x) = s_0 + s_1 x + s_2 x^2 + \cdots + s_{N-1} x^{N-1}$. If s is a periodic sequence with the first period $s^N$, then,

$$s(x) = s^N(x)(1 + x^N + x^{2N} + \cdots) = \frac{s^N(x)}{1-x^N} = \frac{s^N(x)/\gcd(s^N(x), 1-x^N)}{(1-x^N)/\gcd(s^N(x), 1-x^N)} = \frac{g(x)}{f_s(x)}$$

where $f_s(x) = (1-x^N)/\gcd(s^N(x), 1-x^N)$, $g(x) = s^N(x)/\gcd(s^N(x), 1-x^N)$

　　Obviously, $\gcd(g(x), f_s(x))=1$, $\deg(g(x)) < \deg(f_s(x))$, $f_s(x)$ is the minimal polynomial of s, and the degree of $f_s(x)$ is the linear complexity of s, that is $\deg(f_s(x))=c(s)$[2].

　　Let us recall some results in finite field theory[7] and number theory[8].

　　**Definition 1.1** Let n be a positive integer. The polynomial $\Phi_n(x) = \prod_{0<j<n, (j,n)=1}(x - X_n^j)$, where $X_n$ is a n-th primitive unit root and (j,n)=1 denotes j is relatively prime to n, is called the n-th cyclotomic polynomial.

　　**Lemma 1.1** Let p be a prime. Then $\varphi(p^n) = p^n - p^{n-1}$, where n is a positive integer, $\varphi$ is the Euler function.

　　**Lemma 1.2** Let $\Phi_n(x)$ be the n-th cyclotomic polynomial. Then $\Phi_n(x)$ is irreducible over GF(q) if and only if that q is a primitive root modulo n, that is the order of q modulo n is $\varphi(n)$.

　　**Lemma 1.3** Let p be a prime and m a positive integer. Then $\Phi_{p^m}(x) = \Phi_p(x^{p^{m-1}})$.

　　**Proof:** Since p is a prime,

$$\Phi_p(y) = \prod_{0<j<p, (j,p)=1}(y - X_p^j) = \prod_{0 \leq j < p}(y - X_p^j)/(y - X_p^0) = \frac{y^p - 1}{y - 1} = 1 + y + y^2 + \cdots + y^{p-1}.$$

　　Note that $X_{p^m}^j = \exp(\frac{2\pi i}{p^m} j) = \exp(\frac{2\pi i}{p^{m-1}} k) = X_{p^{m-1}}^k$, where j=pk,

$$\Phi_{p^m}(x) = \frac{x^{p^m}-1}{\prod_{0 \leq j < p^m, p|j}(x - x_{p^m}^j)} = \frac{x^{p^m}-1}{\prod_{0 \leq k < p^{m-1}}(x - x_{p^{m-1}}^k)} = \frac{x^{p^m}-1}{x^{p^{m-1}}-1} = \Phi_p(x^{p^{m-1}}). \blacksquare$$

**Lemma 1.4** Let p and q be prime numbers, m and n be positive integers. Let $\Phi_{p^m}(x)^{q^n}$ denote $[\Phi_{p^m}(x)]^{q^n}$. Then $\Phi_{p^m}(x)^{q^n} = \Phi_{p^m}(x^{q^n}) = \Phi_p(x^{q^n p^{m-1}})$, where the operation is over GF(q).

**Proof:** Since the operation is over GF(q), so qy=0.

As q is prime, thus $q | \binom{q}{i}$, $0 < i < q$, $\therefore (a+b)^q = a^q + b^q$,

$\therefore \Phi_p(y)^q = (1 + y + y^2 + \cdots + y^{p-1})^q = 1 + y^q + y^{2q} + \cdots + y^{(p-1)q} = \Phi_p(y^q)$,

By analogy, $\Phi_p(y)^{q^n} = \Phi_p(y^{q^n})$.

By lemma 1.3, $\Phi_{p^m}(x)^{q^n} = \Phi_p(x^{p^{m-1}})^{q^n} = \Phi_p(x^{p^{m-1} q^n}) = \Phi_p[(x^{q^n})^{p^{m-1}}] = \Phi_{p^m}(x^{q^n}). \blacksquare$

**Lemma 1.5** Let p be a prime, q a prime and a primitive root modulo $p^2$. Then q is a primitive root modulo $p^n$, $n \geq 1$, so $\Phi_{p^n}(x)$ is irreducible over GF(q).

**Proof:** We first prove that q is a primitive root modulo p.

Suppose that $q^{\frac{p-1}{m}} \equiv 1 \pmod{p}$, where m is a positive integer, then $q^{\frac{p-1}{m}} = 1 + kp$, where k is a positive integer,

$\therefore q^{\frac{p(p-1)}{m}} = (q^{\frac{p-1}{m}})^p = (1+kp)^p \equiv 1 \pmod{p^2}$.

Since q is a primitive root modulo $p^2$, so m must be 1, thus q is a primitive root modulo p.

Secondly, we prove that q is a primitive root modulo $p^n$, $n \geq 2$, by induction,
Suppose the claim is true for $n \geq 2$, we now consider the case n+1.

Obviously, $q^{p^{n-1} - p^{n-2}} = 1 + kp^{n-1}$, where k is a positive integer;

Since q is a primitive root modulo $p^n$, so k is not divisible by p.

$q^{p^n - p^{n-1}} = (q^{p^{n-1} - p^{n-2}})^p = 1 + kp^n + rp^{n+1}$, where r is a positive integer,

$\therefore q^{p^n - p^{n-1}} \neq 1 + sp^{n+1}$, where s is a positive integer.

Suppose the order of q modulo $p^{n+1}$ is t, then $q^t = 1 + mp^{n+1}$, where m is a positive integer, and $t | j(p^{n+1})$.

By lemma 1.1, $t | (p^{n+1} - p^n)$.

As $q^t = 1 + (mp)p^n$, so $(p^n - p^{n-1}) | t$, thus $t = p^n - p^{n-1}$ or $p^{n+1} - p^n$.

Since $t = p^n - p^{n-1}$ contradicts the fact that $q^{p^n - p^{n-1}} \neq 1 + sp^{n+1}$, thus $t = p^{n+1} - p^n$.

Therefore, q is a primitive root modulo $p^n$, $n \geq 1$.

By lemma 1.2, $\Phi_{p^n}(x)$ is irreducible over GF(q). $\blacksquare$

## 2. Main theorems concerning algorithms

The following lemma and its proof is from [4].

**Lemma 2.1** Let $a = (a_0, a_1, \cdots, a_{N-1})$ be a finite sequence over GF(q), where $N = p^m$, p is a prime and q is a primitive root modulo $p^2$. Let us denote a(x) as the generated function of the finite sequence $(a_0, a_1, \cdots, a_{N-1})$ and $A_i = (a_{(i-1)p^{m-1}}, a_{(i-1)p^{m-1}+1}, \cdots, a_{ip^{m-1}-1})$, $i = 1, 2, \cdots, p$. Then $(\Phi_{p^m}(x), a(x)) \neq 1$, that is $\Phi_{p^m}(x) | a(x)$ if and only if $A_1 = A_2 = \cdots = A_p$.

**Proof:** Let $A_i(x)$ denote the generated function of $A_i$, $i = 1, 2, \cdots, p$. Then,

$a(x) = A_1(x) + x^{p^{m-1}} A_2(x) + \cdots + x^{(p-1)p^{m-1}} A(x)_p$.

We first show the necessity.

As $\Phi_{p^m}(x) | a(x)$, let $a(x) = t(x) \Phi_{p^m}(x)$, where t(x) is a polynomial over GF(q).

From lemma 1.3, $\Phi_{p^m}(x) = 1 + x^{p^{m-1}} + \cdots + x^{(p-1)p^{m-1}}$,

$\therefore A_1(x) - t(x) + x^{p^{m-1}}(A_2(x) - t(x)) + \cdots + x^{(p-1)p^{m-1}}(A(x)_p - t(x)) = 0$.

Since $\deg(a(x)) < p^m$ and $\deg(\Phi_{p^m}(x)) = (p-1)p^{m-1}$, so $\deg(t(x)) < p^m - (p-1)p^{m-1} = p^{m-1}$;

As $\deg(A_i(x)) < p^{m-1}$, so $\deg(A_i(x) - t(x)) < p^{m-1}$, thus $A_i(x) - t(x) = 0$, $i = 1, 2, \cdots, p$, hence $A_1 = A_2 = \cdots = A_p$.

Now come to the sufficiency.

As $A_1 = A_2 = \cdots = A_p$, so $a(x) = A_1(x)(1 + x^{p^{m-1}} + \cdots + x^{(p-1)p^{m-1}}) = A_1(x)\Phi_{p^m}(x)$, hence $\Phi_{p^m}(x) | a(x)$. ∎

**Theorem 2.1**. Let $a = (a_0, a_1, \cdots, a_{N-1})$ be a finite sequence over $GF(q)$, where $N = q^n p^m$, $p$ and $q$ are prime numbers, $q$ is a primitive root modulo $p^2$. Let us denote $a(x)$ as the generated function of the finite sequence $(a_0, a_1, \cdots, a_{N-1})$, $M = q^{n-1}p^{m-1}$, $A_i = (a_{(i-1)M}, a_{(i-1)M+1}, \cdots, a_{iM-1})$, $A_i(x)$ be the generated function of $A_i$, $i = 1, 2, \cdots, qp$. Then,

(i).   $\Phi_{p^m}(x)^{q^{n-1}} | a(x)$ if and only if $A_1 + A_{p+1} + \cdots + A_{(q-1)p+1} = A_2 + A_{p+2} + \cdots + A_{(q-1)p+2} = \cdots = A_p + A_{2p} + \cdots + A_{qp}$;

(ii).  If $\Phi_{p^m}(x)^{q^{n-1}} | a(x)$, then $\gcd(a(x), \Phi_{p^m}(x)^{q^n}) = \Phi_{p^m}(x)^{q^{n-1}} \gcd(a'(x), \Phi_{p^m}(x)^{(q-1)q^{n-1}})$, where $a'(x) = A'_1(x) + A'_2(x) x^M + \cdots + A'_{qp}(x) x^{(qp-1)M}$, $A'_1 = A_1$, $A'_2 = -A_1 + A_2$, $\cdots$, $A'_{qp} = -A_{p-1} - A_{2p-1} - \cdots - A_{qp-1} + A_p + A_{2p} + \cdots + A_{qp}$;

(iii). If $\Phi_{p^m}(x)^{q^{n-1}} | a(x)$ not true, then $\gcd(a(x), \Phi_{p^m}(x)^{q^n}) = \gcd(a(x), \Phi_{p^m}(x)^{q^{n-1}}) = \gcd(\sum_{i=1}^{p}[A_i(x) + A_{p+i}(x) + \cdots + A_{(q-1)p+i}(x)]x^{(i-1)M}, \Phi_{p^m}(x)^{q^{n-1}})$

(iv).  $\gcd(a(x), 1-x^{qM}) = \gcd(\sum_{i=1}^{q}[A_i(x) + A_{q+i}(x) + \cdots + A_{(p-1)q+i}(x)]x^{(i-1)M}, 1-x^{qM})$

(v).   $\gcd(a(x), 1-x^N) = \gcd(a(x), 1-x^{qM}) \gcd(a(x), \Phi_{p^m}(x)^{q^n})$

**Proof:** (i) As $q$ is a primitive root modulo $p^2$, we know that $\Phi_{p^m}(x)$ is irreducible over $GF(q)$.

From lemma 1.4, $\Phi_{p^m}(x)^{q^{n-1}} = 1 + x^M + \cdots + x^{(p-1)M}$, let $b = \Phi_{p^m}(x)^{q^{n-1}}$. Then,

$a(x) = A_1(x) + A_2(x) x^M + \cdots + A_{qp}(x) x^{(qp-1)M}$

$= A'_1(x)b + A'_2(x)bx^M + \cdots + A'_p(x)bx^{(p-1)M}$

$+ A'_{p+1}(x)bx^{pM} + A'_{p+2}(x)bx^{(p+1)M} + \cdots + A'_{2p}(x)bx^{(2p-1)M}$

$+ \cdots\cdots$

$+ A'_{(q-1)p+1}(x)bx^{(q-1)pM} + A'_{(q-1)p+2}(x)bx^{[(q-1)p+1]M} + \cdots + A'_{qp}(x)bx^{(qp-1)M}$

$- [A_p(x) + A_{2p}(x) + \cdots + A_{qp}(x)]bx^{qpM}$

$+ \{[A_1(x) + A_{p+1}(x) + \cdots + A_{(q-1)p+1}(x)] + [A_2(x) + A_{p+2}(x) + \cdots + A_{(q-1)p+2}(x)]x^M + \cdots + [A_p(x) + A_{2p}(x) + \cdots + A_{qp}(x)]x^{(p-1)M}\}x^{qpM}$       (1)

where, $A'_1 = A_1$, $A'_2 = -A_1 + A_2$, $\cdots$, $A'_p = -A_{p-1} + A_p$,

$A'_{p+1} = -A_p + A_1 + A_{p+1}$, $A'_{p+2} = -A_1 - A_{p+1} + A_2 + A_{p+2}$, $\cdots$, $A'_{2p} = -A_{p-1} - A_{2p-1} + A_p + A_{2p}$,

$\cdots\cdots$,

$A'_{(q-1)p+1} = -A_p - A_{2p} - \cdots - A_{(q-1)p} + A_1 + A_{p+1} + \cdots + A_{(q-1)p+1}$,

$A'_{(q-1)p+2} = -A_1 - A_{p+1} - \cdots - A_{(q-1)p+1} + A_2 + A_{p+2} + \cdots + A_{(q-1)p+2}$, $\cdots$,

$A'_{qp} = -A_{p-1} - A_{2p-1} - \cdots - A_{qp-1} + A_p + A_{2p} + \cdots + A_{qp}$.

To understand equality (1), we can add all the items of the right side concerning one polynomial, such as $A_2(x)$.

$x^M[A_2(x)b - A_2(x)bx^M + A_2(x)bx^{pM} - A_2(x)bx^{(p+1)M} + \cdots + A_2(x)bx^{(q-1)pM} - A_2(x)bx^{[(q-1)p+1]M} + A_2(x)x^{qpM}]$

$= x^M A_2(x)[b(1 + x^{pM} + \cdots + x^{(q-1)pM}) - bx^M(1 + x^{pM} + \cdots + x^{(q-1)pM}) + x^{qpM}]$

$= x^M A_2(x)[b(1 - x^M)(1 + x^{pM} + \cdots + x^{(q-1)pM}) + x^{qpM}]$

$= x^M A_2(x)[(1 - x^{pM})(1 + x^{pM} + \cdots + x^{(q-1)pM}) + x^{qpM}] = A_2(x) x^M (1 - x^{qpM} + x^{qpM}) = x^M A_2(x)$.

In the case of $A_{qp}(x)$,

$x^{(qp-1)M}(A_{qp}(x)b - A_{qp}(x)bx^M + A_{qp}(x)x^{pM}) = A_{qp}(x)x^{(qp-1)M}[b(1-x^M) + x^{pM}] = A_{qp}(x)x^{(qp-1)M}$.

By analogy, we can verify other items of equality (1).

Since $(\Phi_{p^m}(x)^{q^{n-1}}, x^{qpM}) = 1$, so $\Phi_{p^m}(x)^{q^{n-1}} | a(x) \Leftrightarrow$

$\Phi_{p^m}(x)^{q^{n-1}} | \{[A_1(x) + A_{p+1}(x) + \cdots + A_{(q-1)p+1}(x)] + [A_2(x) + A_{p+2}(x) + \cdots + A_{(q-1)p+2}(x)]x^M + \cdots + [A_p(x) + A_{2p}(x) + \cdots + A_{qp}(x)]x^{(p-1)M}\}$.

From lemma 2.1, we have $\Phi_{p^m}(x)^{q^{n-1}} | a(x) \Leftrightarrow$

$A_1(x) + A_{p+1}(x) + \cdots + A_{(q-1)p+1}(x) = A_2(x) + A_{p+2}(x) + \cdots + A_{(q-1)p+2}(x) = \cdots = A_p(x) + A_{2p}(x) + \cdots + A_{qp}(x)$

$\Leftrightarrow A_1 + A_{p+1} + \cdots + A_{(q-1)p+1} = A_2 + A_{p+2} + \cdots + A_{(q-1)p+2} = \cdots = A_p + A_{2p} + \cdots + A_{qp}$.

(ii) If $\Phi_{p^m}(x)^{q^{n-1}} | a(x)$, then $A_1 + A_{p+1} + \cdots + A_{(q-1)p+1} = A_2 + A_{p+2} + \cdots + A_{(q-1)p+2} = \cdots = A_p + A_{2p} + \cdots + A_{qp}$.

Let $a'(x) = a(x)/\Phi_{p^m}(x)^{q^{n-1}}$. From equality (1), $a'(x) = A'_1(x) + A'_2(x)x^M + \cdots + A'_{qp}(x)x^{(qp-1)M}$.

Thus, $\gcd(a(x), \Phi_{p^m}(x)^{q^n}) = \Phi_{p^m}(x)^{q^{n-1}} \gcd(a'(x), \Phi_{p^m}(x)^{(q-1)q^{n-1}})$.

(iii) If $\Phi_{p^m}(x)^{q^{n-1}} | a(x)$ not true, then $\gcd(a(x), \Phi_{p^m}(x)^{q^n}) = \gcd(a(x), \Phi_{p^m}(x)^{q^{n-1}})$.

From equality (1), $\gcd(a(x), \Phi_{p^m}(x)^{q^n}) = \gcd(\sum_{i=1}^{p}[A_i(x) + A_{p+i}(x) + \cdots + A_{(q-1)p+i}(x)]x^{(i-1)M}, \Phi_{p^m}(x)^{q^{n-1}})$.

(iv) $a(x) = A_1(x) + A_2(x)x^M + \cdots + A_{qp}(x)x^{(qp-1)M}$

$= \{A_1(x) + [A_1(x) + A_{q+1}(x)]x^{qM} + \cdots + [A_1(x) + A_{q+1}(x) + \cdots + A_{qp-2q+1}(x)]x^{(qp-2q)M}\}(1-x^{qM})$

$+ [A_1(x) + A_{q+1}(x) + \cdots + A_{qp-2q+1}(x) + A_{qp-q+1}(x)]x^{(qp-q)M}$

$+ \cdots\cdots +$

$\{A_q(x) + [A_q(x) + A_{2q}(x)]x^{qM} + \cdots + [A_q(x) + A_{2q}(x) + \cdots + A_{qp-q}(x)]x^{(qp-2q)M}\}x^{(q-1)M}(1-x^{qM})$

$+ [A_q(x) + A_{2q}(x) + \cdots + A_{qp-q}(x) + A_{qp}(x)]x^{(qp-1)M}$,

Since $\gcd(x^{(qp-q)M}, 1-x^{qM}) = 1$, thus,

$\gcd(a(x), 1-x^{qM}) = \gcd([A_1(x) + A_{q+1}(x) + \cdots + A_{qp-2q+1}(x) + A_{qp-q+1}(x)] + \cdots + [A_q(x) + A_{2q}(x) + \cdots + A_{qp-q}(x) + A_{qp}(x)]x^{(q-1)M}, 1-x^{qM})$

(v) Since $\dfrac{1-x^N}{1-x^{qM}} = 1 + x^{qM} + \cdots + x^{q(p-1)M} = \Phi_{p^m}(x^q)^{q^{n-1}} = \Phi_{p^m}(x)^{q^n}$ and $\gcd(1-x^{qM}, \Phi_{p^m}(x)^{q^n}) = 1$, thus,

$\gcd(a(x), 1-x^N) = \gcd(a(x), 1-x^{qM}) \gcd(a(x), \Phi_{p^m}(x)^{q^n})$. ∎

**Theorem 2.2**. Let s be a sequence over GF(q) with period N and $a = (a_0, a_1, \cdots, a_{N-1})$ the first period, where $N = q^n p^m$, p and q are prime numbers, q is a primitive root modulo $p^2$. Let us denote $a(x)$ as the generated function of the finite sequence $(a_0, a_1, \cdots, a_{N-1})$, $M = q^{n-1}p^{m-1}$, $A_i = (a_{(i-1)M}, a_{(i-1)M+1}, \cdots, a_{iM-1})$, $A_i(x)$ be the generated function of $A_i$, $i = 1, 2, \cdots, qp$. Then $f_s(x) = f_{(b)}(x) \cdot \Phi_{p^m}(x)^z$, hence $c(s) = c((b)) + (p-1)p^{m-1}z$, where $b = (A_1 + A_{q+1} + \cdots + A_{qp-q+1}, \cdots, A_q + A_{2q} + \cdots + A_{qp})$, $(b)$ denotes the sequence with the first period b; $\deg(\Phi_{p^m}(x)) = j(p^m) = (p-1)p^{m-1}$, $\Phi_{p^m}(x)^z = \Phi_{p^m}(x)^{q^n}/\gcd(\Phi_{p^m}(x)^{q^n}, a(x))$, hence $z = q^n - t$, where t is the power exponent of $\Phi_{p^m}(x)$ in $\gcd(\Phi_{p^m}(x)^{q^n}, a(x))$.

**Proof:** From (iv) and (v) of theorem 2.1,

$f_s(x) = (1-x^N)/\gcd(a(x), 1-x^N) = [(1-x^{qM})/\gcd(a(x), 1-x^{qM})] \cdot [\Phi_{p^m}(x)^{q^n}/\gcd(a(x), \Phi_{p^m}(x)^{q^n})]$

$= [(1-x^{qM})/\gcd([A_1(x) + A_{q+1}(x) + \cdots + A_{qp-2q+1}(x) + A_{qp-q+1}(x)] + \cdots + [A_q(x) + A_{2q}(x) + \cdots + A_{qp-q}(x) + A_{qp}(x)]x^{(q-1)M}, 1-x^{qM})] \cdot [\Phi_{p^m}(x)^{q^n}/\gcd(a(x), \Phi_{p^m}(x)^{q^n})]$

$= f_{(b)}(x) \cdot \Phi_{p^m}(x)^z$, where (b) with the period qM.

It is easy to show that $c(s)=c((b))+(p-1)p^{m-1}z$. ∎

## 3. Algorithms to compute the linear complexity of sequences over GF(q) with period $q^n p^m$

The following algorithm was presented in [4] as algorithm 1.

**Algorithm 3.1** Let s be a sequence over GF(q) with period $N=p^n$ and the first period be denoted as $a=(a_0, a_1, \cdots, a_{N-1})$, where q is a primitive root modulo $p^2$.

Initial values: $a=(a_0, a_1, \cdots, a_{N-1})$ is the first period of s, $k=p^n$, $c=0$, $f=1$.

(i) If $a=(0,\cdots,0)$, then end; if $k=1$, then $c=c+1, f=(1-x)f$, end

(ii) $k=k/p$, let $A_i = (a_{(i-1)k}, a_{(i-1)k+1}, \cdots, a_{ik-1})$, $i=1,2,\cdots,p$;

(iii) If $A_1=A_2=\cdots=A_p$, then $a=A_1$; if $A_1=A_2=\cdots=A_p$ not true, then $a=A_1+A_2+\cdots+A_p$, $c=c+(p-1)k$, $f=f\Phi_{pk}(x)$;

(iv) go to (i).

(v) The final. The linear complexity of s: $c(s)=c$; the minimal polynomial of s: $f_s(x)=f$.

From lemma 2.1, it is easy to show the correctness of the algorithm and it computes the minimal polynomial in (n+1) loops at most. The reader is referred to [4] for a detailed proof.

**Lemma 3.1** Let s be a sequence over GF(q) with period $N=q^n$ and $a=(a_0, a_1, \cdots, a_{N-1})$ be the first period, where q is a prime. Let us denote $a(x)$ as the generated function of the finite sequence a, $M=q^{n-1}$, $A_i = (a_{(i-1)M}, a_{(i-1)M+1}, \cdots, a_{iM-1})$, $A_i(x)$ be the generated function of $A_i$, $i=1,2,\cdots,q$. Then,

(i) $(1-x^M)|a(x)$ if and only if $\sum_{i=1}^{q} A_i = 0$;

(ii) if $(1-x^M)|a(x)$, then $\dfrac{a(x)}{1-x^M} = A_1(x)+[A_1(x)+A_2(x)]x^M+\cdots+[A_1(x)+A_2(x)+\cdots+A_q(x)]x^{(q-1)M}$;

(iii) if $(1-x^M)|a(x)$ not true, then $\gcd(a(x), 1-x^N) = \gcd(A_1(x)+A_2(x)+\cdots+A_q(x), 1-x^M)$

(iv) if $(1-x^M)|a(x)$, then $\gcd(a(x),1-x^N) = (1-x^M)\gcd(A_1(x)+[A_1(x)+A_2(x)]x^M+\cdots+[A_1(x)+A_2(x)+\cdots+A_q(x)]x^{(q-1)M}, (1-x^M)^{q-1})$.

**Proof:** From the following equality, we have (i) and (ii),

$a(x)= A_1(x)+A_2(x)x^M+\cdots+ A_q(x)x^{(q-1)M}$

$=(1-x^M)\{A_1(x)+[A_1(x)+A_2(x)]x^M+\cdots+[A_1(x)+A_2(x)+\cdots+A_q(x)]x^{(q-1)M}\}+[A_1(x)+A_2(x)+\cdots+A_q(x)]x^{qM}$.

Note that the operation is over GF(q) and q is a prime,

$(1-x)^q = 1-x^q$,

$(1-x)^{q^2} = (1-x^q)^q = 1-x^{q^2}$,

By analogy, $(1-x)^N = (1-x)^{q^n} = 1-x^{q^n} = 1-x^N$,

∴ $1-x^N = (1-x^M)^q$

Therefore, we have (iii) and (iv). ∎

From lemma 3.1, it is easy to show the correctness of the following algorithm.

**Algorithm 3.2** Let s be a sequence over GF(q) with period $N=q^n$ and $a=(a_0, a_1, \cdots, a_{N-1})$ be the first period, where q is a prime.

Initial values: $a=s^N$, $l=q^n$, $c=0$, $f=1$.

(i) If $l=1$, then {If $a=(0)$, then end; else $c=c+1, f=(1-x)f$, end.}

(ii) If $l \neq 1$, then $l=l/q$, denote $M=l$, $A_i = (a_{(i-1)M}, a_{(i-1)M+1}, \cdots, a_{iM-1})$, $i=1,2,\cdots,q$. Set count=0.

(iii) If $A_1+A_2+\cdots+A_q = 0$, then {count= count+1, if count<q, then set $A_i = A_i + A_{i-1}$ ($i=2,\cdots,q$, sequentially), repeat (iii); if count=q, then end.}

(iv) If $A_1+A_2+\cdots+A_q \neq 0$, then $a= A_1+A_2+\cdots+A_q$, $c=c+(q-count-1)M$, $f=f(1-x^M)^{q-count-1}$, go to (i).

(v) The final. The linear complexity of s: $c(s)=c$; the minimal polynomial of s: $f_s(x)=f$.

The algorithm computes the minimal polynomial in n(q-1)+1 loops at most.

Let s be a sequence over GF(q) with period N and a=($a_0, a_1, \cdots, a_{N-1}$) the first period, where N= $q^n p^m$, p is a prime, q is a prime and a primitive root modulo $p^2$. From theorem 2.2, the computation of $f_s(x)$ is equivalent to that of $f_{(b)}(x)$ and $\Phi_{p^m}(x)^z$, so we first introduce an algorithm to compute $\Phi_{p^m}(x)^z$.

**Algorithm 3.3**. Initial values: a=($a_0, a_1, \cdots, a_{N-1}$) is the first period of s, $l=q^n$, c=0, f=1, we denote k=$p^{m-1}$.

(i) If a=(0, $\cdots$,0), then end; if $l=1$, then{let $A_i$ = ($a_{(i-1)k}$, $a_{(i-1)k+1}$, $\cdots$, $a_{ik-1}$), i=1,2,$\cdots$,p. If $A_1 = A_2 = \cdots = A_p$, then end; else, c=c+(p-1)k, f=f$\Phi_{pk}(x)$, end.}

(ii) If $l \neq 1$, then $l=l/q$, denote M=$l$k, $A_i$ = ($a_{(i-1)M}$, $a_{(i-1)M+1}$, $\cdots$, $a_{iM-1}$), i=1,2,$\cdots$,qp. Set count=0.

(iii) If $A_1 + A_{p+1} + \cdots + A_{(q-1)p+1} = A_2 + A_{p+2} + \cdots + A_{(q-1)p+2} = \cdots = A_p + A_{2p} + \cdots + A_{qp}$, then{count= count+1, if count<q, set $A'_1, A'_2, \cdots, A'_{qp}$ according to the definition in theorem 2.1, set $A_1 = A'_1, A_2 = A'_2, \cdots, A_{qp} = A'_{qp}$, repeat (iii); if count=q, end. }

(iv) If $A_1 + A_{p+1} + \cdots + A_{(q-1)p+1} = A_2 + A_{p+2} + \cdots + A_{(q-1)p+2} = \cdots = A_p + A_{2p} + \cdots + A_{qp}$ not true, a=( $A_1 + A_{p+1} + \cdots + A_{(q-1)p+1}$, $A_2 + A_{p+2} + \cdots + A_{(q-1)p+2}, \cdots, A_p + A_{2p} + \cdots + A_{qp}$), c=c+(q-count-1)(p-1)M, f=f$\Phi_{pk}(x)^{(q-count-1)l}$, go to (i)

(v) The final. f is $\Phi_{p^m}(x)^z$, c is (p-1)$p^{m-1}$z.

From theorem 2.1 and theorem 2.2, algorithm 3.3 immediately follows. It computes $\Phi_{p^m}(x)^z$ in n(q-1)+1 loops at most.

Now come to our main result, an efficient algorithm for computing the linear complexity and the minimal polynomial of sequences over GF(q).

**Algorithm 3.4**. Initial values: a=($a_0, a_1, \cdots, a_{N-1}$) is the first period of s, $l= q^n$, k=$p^m$, c=0, f=1.

(i) If a=(0, $\cdots$,0), then end; if k>1, then k=k/p and go to (iv).

(ii) If $l=1$, then c=c+1,f=(1-x)f, end; else if $l \neq 1$, then $l=l/q$, denote M=$l$, $A_i$ = ($a_{(i-1)M}$, $a_{(i-1)M+1}$, $\cdots$, $a_{iM-1}$), i=1,2,$\cdots$,q. Set count=0.

(iii) If $A_1 + A_2 + \cdots + A_q = 0$, then {count= count+1, if count<q, then set $A_i = A_i + A_{i-1}$ (i=2,$\cdots$,q, sequentially), repeat (iii); if count=q, then end.} else if $A_1 + A_2 + \cdots + A_q \neq 0$, then a= $A_1 + A_2 + \cdots + A_q$, c=c+(q-count-1)M, f=f$(1-x^M)^{q-count-1}$, go to (i).

(iv) If $l = 1$, then{let $A_i$ = ($a_{(i-1)k}$, $a_{(i-1)k+1}$, $\cdots$, $a_{ik-1}$), i=1,2,$\cdots$,p. If $A_1 = A_2 = \cdots = A_p$, then a= $A_1$, go to (i); else a= $A_1 + A_2 + \cdots + A_p$, c=c+(p-1)k, f=f$\Phi_{pk}(x)$, go to (i).}

(v) If $l \neq 1$, then $l=l/q$, let M=$l$k, $A_i$ = ($a_{(i-1)M}$, $a_{(i-1)M+1}$, $\cdots$, $a_{iM-1}$), i=1,2,$\cdots$,qp. b=($A_1 + A_{q+1} + \cdots + A_{qp-2q+1} + A_{qp-q+1}, \cdots, A_q + A_{2q} + \cdots + A_{qp-q} + A_{qp}$). Set count=0.

(vi) If $A_1 + A_{p+1} + \cdots + A_{(q-1)p+1} = A_2 + A_{p+2} + \cdots + A_{(q-1)p+2} = \cdots = A_p + A_{2p} + \cdots + A_{qp}$, count= count+1, if count<q, set $A'_1, A'_2, \cdots, A'_{qp}$ according to the definition in theorem 2.1, set $A_1 = A'_1, A_2 = A'_2, \cdots, A_{qp} = A'_{qp}$, repeat (vi); if count=q, then a=b, $l=q^n$, go to (i).

(vii) If $A_1 + A_{p+1} + \cdots + A_{(q-1)p+1} = A_2 + A_{p+2} + \cdots + A_{(q-1)p+2} = \cdots = A_p + A_{2p} + \cdots + A_{qp}$ not true, a=( $A_1 + A_{p+1} + \cdots + A_{(q-1)p+1}$, $A_2 + A_{p+2} + \cdots + A_{(q-1)p+2}, \cdots, A_p + A_{2p} + \cdots + A_{qp}$), c=c+(q-count-1)(p-1)M, f=f$\Phi_{pk}(x)^{(q-count-1)l}$.

(viii) If $l=1$, then{let $A_i$ = ($a_{(i-1)k}$, $a_{(i-1)k+1}$, $\cdots$, $a_{ik-1}$), i=1,2,$\cdots$,p. If $A_1 = A_2 = \cdots = A_p$, then a=b, $l=q^n$, go to (i); else, c=c+(p-1)k, f=f$\Phi_{pk}(x)$, a=b, $l=q^n$, go to (i).}

(ix) If $l \neq 1$, then $l=l/q$, let M=$l$k, $A_i$ = ($a_{(i-1)M}$, $a_{(i-1)M+1}$, $\cdots$, $a_{iM-1}$), i=1,2,$\cdots$,qp. Set count=0, go to (vi).

(x) The final. The linear complexity of s: c(s)=c; the minimal polynomial of s: $f_s(x)$=f.

From theorem 2.1, theorem 2.2, algorithm 3.1, algorithm 3.2 and algorithm 3.3, we know that algorithm 3.4 is correct. With a similar argument as that of algorithm 3.2 in [5], it is easy to show that it computes the minimal polynomial in [n(q-1)+1](m+1) loops at most.